\documentstyle[preprint,aps]{revtex}
\begin{document}
\draft
\title{Two definitions of the electric polarizability of a bound system \\ 
in relativistic quantum theory}
\author{F.A.B. Coutinho$^a$, Y. Nogami$^b$ and Lauro Tomio$^{b,c}$}
\address{
$^a$ Faculdade de Medicina, Universidade de S\~ao Paulo,  
01246-903, S\~ao Paulo, Brazil \\
$^b$ Department of Physics and Astronomy, McMaster University,\\  
Hamilton, Ontario, Canada L8S 4M1\\
$^c$ Instituto de F\'{\i}sica Te\'orica, Universidade Estadual Paulista,
01405-900, S\~ao Paulo, Brazil
}
\date{\today}
\maketitle
%\pacs{PACS 03.65.-w, 12.39.Ba, 12.39.Ki}
\begin{abstract}
For the electric polarizability of a bound system in relativistic 
quantum theory, there are two definitions that have appeared in the 
literature. They differ depending on whether or not the vacuum 
background is included in the system. A recent confusion in this 
connection is clarified.
\end{abstract}
\vskip 1cm

Recently three papers appeared in this journal discussing the electric 
polarizability (EP) of a relativistic bound system \cite{1,2,3}. In 
Refs. \cite{1,2,3} it was illustrated by model calculations that the EP 
of a relativistic system can be negative when the interaction that 
binds the system is very strong. The model used in Refs. \cite{1,2,3} 
is a particle that is bound in a given potential in one dimension and 
subject to the Dirac equation. Reference \cite{3} presents interesting 
discussions on effects of the vacuum background on the EP on the basis 
of Dirac's hole theory (HT).

The purpose of this note is to point out that the definition of the EP 
that was assumed in Refs. \cite{1,2,3} is different from the one that 
was used in earlier papers \cite{4,5}. The two definitions differ 
depending on whether or not the vacuum background is regarded as part 
of the system.  We are not going to argue that one is correct and the 
other is wrong but we have to be clear about the distinction between 
the two. Unfortunately the two definitions are apparently confused in 
Ref. \cite{3}; see the remark at the end of this note. As we emphasize 
below, if the vacuum background is included, the EP is positive no 
matter how strong the binding interaction is. 

Consider a bound system like the hydrogen atom. When it is perturbed 
by an external electric field, the system is polarized and its energy 
shifts. Assume that the electric field ${\bf E}$ is constant and weak. 
Then the energy shift $W$ takes the form
\begin{equation}
W = -\frac{1}{2}\alpha{\bf E}^2.
\label{1}
\end{equation}
This $W$ is nothing but the second order energy shift caused by the 
perturbation due to ${\bf E}$. The coefficient $\alpha$ is the EP of 
the system. This is how the EP is defined but there can be different 
definitions depending on how the system is interpreted. In Ref. 
\cite{1,2,3} the bound system was regarded as a single particle 
system, a particle bound in a given potential. In quantum field 
theory (QFT) or equivalently in HT, in addition to the bound 
particle, the vacuum background is considered. When the vacuum 
background is interpreted as an integral part of the bound system, 
it is no longer a single particle system. This is how the bound 
system is treated in Refs. \cite{4,5}.

Let us elaborate on the two definitions. As a way of setting up 
notation, let us start with the problem as that of the single-particle 
quantum mechanics. Let the Hamiltonian be
\begin{equation}
H = H_0 + V\, ,
\label{2}
\end{equation}
where $H_0$ is the Dirac Hamiltonian with a binding potential and $V$
is the external perturbation. More explicitly, $V=-q{\bf E}\cdot 
{\bf r}$ where $q$ is the charge of the particle. We take $H_0$ as 
the unperturbed Hamiltonian and treat $V$ by perturbation theory. It 
is understood that the solutions of the Dirac equation with $H_0$ are
known for all stationary states,
\begin{equation}
H_0|i\rangle  = \epsilon_{i}|i\rangle\, ,\hspace{0.3in}
H_0|-j\rangle  = \epsilon_{-j}|-j\rangle\, ,
\label{3}
\end{equation}
where $i=1,2, \cdots$ and $-j=-1,-2, \cdots$. The $|i\rangle $'s 
($|-j\rangle $'s) are positive (negative) energy states; 
$\epsilon_i > 0$ ($\epsilon_{-j} <0$). In particular $|1\rangle
$ is the lowest positive energy state. We are assuming that the
eigenvalues are all discrete but it is straightforward to include
continuum. The $|i\rangle $'s and $|-j\rangle $'s form a complete
orthonormal basis set. For the unperturbed state let us take 
$|1\rangle$, the state of the lowest positive energy \cite{6}. The 
second order energy shift $W_{\rm QM}$ of state $|1\rangle $ caused 
by perturbation $V$ is given by
\begin{equation}
W_{\rm QM} = \sum_{i\neq 1} \frac{|V_{i,1}|^2}
{\epsilon_1 -\epsilon_i} + \sum_{j} \frac{|V_{-j,1}|^2}
{\epsilon_1 -\epsilon_{-j}}\, ,
\label{4}
\end{equation}
where $V_{i,1}\equiv \langle i|V|1\rangle$ and $V_{-j,1}\equiv 
\langle -j|V|1\rangle$.  Suffix QM refers to single-particle quantum 
mechanics. The summation for $i$ ($j$) is for the positive (negative) 
energy intermediate states. The contributions from the negative energy 
intermediate states can make $W_{\rm QM}$ positive \cite{1,2,3}. 

Let us examine the vacuum background following Ref. \cite{3}. In HT 
the vacuum is such that all negative energy states are occupied. We 
replace $W_{\rm QM}$ obtained above with
\begin{equation}
W_1 = \sum_{i\neq 1} \frac{|V_{i,1}|^2}{\epsilon_1 -\epsilon_i}\, ,
\label{5}
\end{equation}
where the Pauli principle excludes the negative energy states as 
intermediate states. On the other hand the vacuum energy also shifts. 
The vacuum energy shift is given by
\begin{equation}
W_{\rm vac} = \sum_{j}W_{-j}\, ,\hspace{0.2in} W_{-j} = \sum_{i\neq
1}\frac{|V_{i,-j}|^2}{\epsilon_{-j} -\epsilon_i}\, .
\label{6}
\end{equation}
Again the summation over $i$ $(j)$ is for positive (negative) energy 
states. The intermediate state of $i=1$ is excluded because it is 
already occupied. If we interpret that the vacuum background is part 
of the system, the total energy shift is given by
\begin{equation}
W_{\rm HT} = W_1 + W_{\rm vac}\, .
\label{7}
\end{equation}
The $W_1$ and $W_{\rm vac}$ are both negative and hence $W_{\rm HT}$ 
is negative.

As shown in Ref. \cite{3} $W_{\rm HT}$ can be rewritten as
\begin{equation}
W_{\rm HT} = W_{\rm QM} + W'_{\rm vac}\, ,
\label{8}
\end{equation}
\begin{equation}
W'_{\rm vac} = \sum_{j}W'_{-j}\, ,\hspace{0.2in} 
W'_{-j} = \sum_{i}\frac{|V_{i,-j}|^2}{\epsilon_{-j} -\epsilon_i}\, .
\label{9}
\end{equation}
The restriction $i\neq 1$ has been removed in the $i$-summation for 
$W'_{-j}$. The $W'_{\rm vac}$ is the vacuum energy shift {\em in the 
absence of the particle in $|1\rangle $}. The $W_{\rm HT}$ of Eq. 
(\ref{8}) contains terms that violate the Pauli principle but such 
terms all cancel out. This is an interesting illustration of 
Feynman's time-honored trick \cite{8}. 

In QFT no negative energy particles appear but antiparticles of 
positive energies appear instead. The unperturbed state that we 
consider is ${c_1}^\dagger |{\rm vac}\rangle$. Here 
$|{\rm vac}\rangle $ is the state that contains no particles nor 
antiparticles at all. The energy of this unperturbed vacuum is zero. 
The ${c_1}^\dagger $ is an operator that creates a particle with 
energy $\epsilon_1$ and wave function associated with $|1\rangle$. 
The $|{\rm vac}\rangle$ and ${c_1}^\dagger |{\rm vac}\rangle$ are 
the ground states of the unperturbed system within the 
zero-particle and one-particle sectors, respectively. Note that 
the particle number is a conserved quantity. The external electric 
field leads to creation of a particle-antiparticle pair, and so on. 
It turns out that HT is equivalent to QFT.

In summary, depending on what we take for the $W$ of Eq. (\ref{1}), 
we have different polarizabilities,
\begin{equation}
W_{\rm QM} = -\frac{1}{2}\alpha_{\rm QM}{\bf E}^2\, ,\hspace{0.3in} 
W_{\rm HT} = W_{\rm QFT} = -\frac{1}{2}\alpha{\bf E}^2\, .
\label{10}
\end{equation}
If we treat the system as a single-particle system, we obtain 
$\alpha_{\rm QM}$ that can be negative as shown in Refs. 
\cite{1,2,3}. The $\alpha_{\rm QM}$ corresponds to $\alpha_{\rm pol}$ 
of Refs. \cite{1,3} and to $P$ of Ref. \cite{2}. If we include the
vacuum background, we obtain $\alpha$ that is related to 
$\alpha_{\rm QM}$ by
\begin{equation}
\alpha = \alpha_{\rm QM} + \alpha '_{\rm vac}\, ,
\label{11}
\end{equation}
where $\alpha '_{\rm vac}$ is the EP of the vacuum in the absence of 
the particle in $|1\rangle$. The $\alpha$ and $\alpha '_{\rm vac}$ 
are respectively equal to $\alpha_1 + \alpha_2$ and $\alpha_3$ of 
Ref. \cite{3}. The $\alpha$ is positive because $W_{\rm HT} = 
W_{\rm QFT}$ is negative as we have discussed.  As an example 
consider the hydrogen atom. The EP of the atom is $\alpha$. The 
$\alpha '_{\rm vac}$ is the EP of the hydrogen ion.

The notion of the EP is important in relation to the London-van der 
Waals force between two neutral atoms, e.g., two hydrogen atoms 
\cite{9}. The inter-atomic force at large distances is proportional 
to $\alpha^2$ \cite{4,9}. This $\alpha$ should be distinguished from
$\alpha_{\rm QM}$. The London-van der Waals force acts between two 
atoms, not just between two bound electrons. The London-van der 
Waals force is the long range part of the two-photon-exchange force 
between atoms. Feinberg et al. \cite{4} developed a dispersion 
theoretical method for the two-photon exchange process. In this 
method relevant matrix elements can be related to the amplitudes 
of Compton scattering from the atoms. The scattering is from the 
entire atoms that include their vacuum background. The EP appears 
in the low-energy limits of the amplitudes. This EP is $\alpha$ 
and not $\alpha_{\rm QM}$ \cite{9}. In explicit calculations of 
the London-van der Waals force, vacuum effects are often ignored. 
This is because the vacuum effects are usually very small. 

Finally let us mention the question raised by Sucher as to the sign 
of the EP \cite{5}. He says that the EP defined in terms of second 
order perturbation theory always gives a positive value (negative 
energy shift) for a system in its ground state. The EP that he
refers to is, in our notation, $\alpha$ and not $\alpha_{\rm QM}$.
He discussed the general validity of this result, for an arbitrary 
elementary system, be it atom, nucleus, or fundamental particle, 
within the framework of relativistic quantum theory. By using 
dispersion theoretical techniques, he examined the Compton 
scattering amplitude of which the low energy limit is related to 
$\alpha$, the EP of the target system. He argued that a possibility 
of negative $\alpha$ may not be excluded as a consequence of only 
the most general principles of relativistic quantum theory. This 
has to do with the high energy limit of the scattering amplitude 
which in turn is related to the ``compositeness" of the target 
system. As far as we know, this question raised by Sucher has not 
been clarified as yet. In discussing the EP in the sense of 
$\alpha_{\rm QM}$, Maize et al. \cite{3} refered to Sucher's question. 
They suggested that the negative $\alpha_{\rm QM}$ that they 
obtained was an answer to Sucher's question. But the EP that Sucher 
examined is, as we said above, $\alpha$ rather than $\alpha_{\rm QM}$.

This work was supported by Funda\c{c}\~{a}o de Amparo \`a Pesquisa do
Estado de S\~{a}o Paulo (FAPESP), Conselho Nacional de Desenvolvimento
Cient\'{\i}fico e Tecnol\'{o}gico (CNPq) and the Natural Sciences and
Engineering Research Council of Canada. LT thanks McMaster University 
for the hospitality he received during his recent visit.


\begin{references}
\bibitem{1} M.A. Maize and C.A. Burkholder, ``Electric polarizability
and the solution of an inhomogeneous differential equation," Am. J.
Phys. {\bf 63}, 244-247 (1995).
\bibitem{2} F.A.B. Coutinho, Y. Nogami and F.M. Toyama, ``Logarithmic
perturbation expansion for the Dirac equation in one dimension:
Application to the polarizability calculation," Am. J. Phys. {\bf 65},
788-794 (1997).
\bibitem{3} M.A.Maize, S. Paulson and A. D'Avanti, ``Electric
polarizability of a relativistic particle," Am. J. Phys. {\bf 65},
888-891 (1997).
\bibitem{4} G. Feinberg, J. Sucher and C.-K. Au, ``The dispersion
theory of dispersion forces," Phys. Reports {\bf 180}, 83-157 (1989);
J. Sucher and G. Feinberg, ``Long-range electromagnetic forces in
quantum theory," in {\em Long-range Casimir Forces}, edited by F.S.
Levin and D.A. Micha (Plenum Press, New York, 1993), Chapt. 5, and
references quoted therein.
\bibitem{5} J. Sucher, ``Sign of the static electric polarizability in
relativistic quantum theory," Phys. Rev. D {\bf 6}, 1798-1800 (1972).
\bibitem{6} We can start with a negative energy bound state. If we do
so, however, we will meet unnecessary, nonessential complications in
the contexts of QFT.
\bibitem{7} This rearrangement of the terms is essentially the same 
as what was done in relating Eqs. (5) and (10) in Ref. \cite{3}.
\bibitem{8} R.P. Feynman, ``The theory of positrons," Phys. Rev. 
{\bf 76}, 749-759 (1949), in particular p. 755.
\bibitem{9} F. London, ``Zur Theorie und Systematik der
Molekularkr\"{a}fte (On the theory and systematics of the molecular
forces)," Zeits. f\"{u}r Physik {\bf 63}, 245-279 (1930); H.G.B.
Casimir and D. Polder, ``The influence of retardation on the London-van
der Waals forces," Phys. Rev. {\bf 73}, 360-372 (1947).

\end{references}
\end{document}